\documentclass[usenatbib]{mn2e}
\usepackage{bm,amssymb}
\usepackage{graphicx}
\usepackage{aas_macros}

\title[Self-Regulated Black Hole Feedback]{Self-Regulated Black Hole
  Growth via Momentum Deposition in Galaxy Merger Simulations}
\author[J. DeBuhr et. al.]{Jackson~DeBuhr,$^{1,2}$ Eliot~Quataert,$^{1,2}$ Chung-Pei~Ma$^2$ and Philip Hopkins$^2$\\
  $^1$Department of Physics, University of California, Berkeley, CA 94720, USA \\
  $^2$Department of Astronomy and Theoretical Astrophysics Center,
  University of California, Berkeley, CA 94720, USA}

\begin{document}

\maketitle

\begin{abstract}
  We perform hydrodynamical simulations of major galaxy mergers using
  new methods for calculating the growth of massive black holes (BH)
  in galactic nuclei and their impact on the surrounding galaxy.  We
  model BH growth by including a subgrid model for accretion produced
  by angular momentum transport on unresolved scales.  The impact of
  the BHs radiation on surrounding gas is approximated by depositing
  momentum into the ambient gas, which produces an outward force away
  from the BH.  We argue that these phenomenological models for BH
  growth and feedback better approximate the interaction between the
  BH and dense gas in galaxies than previous models.  We show that
  this physics leads to self-regulated black hole growth: during the
  peak of activity, the accretion rate onto the BH is largely
  determined by the physics of BH feedback, not the subgrid accretion
  model.  The BH significantly modifies the gas dynamics in the
  galactic nucleus ($\lesssim 300$ pc), but does not generate
  large-scale galactic outflows.  Integrated over an entire galaxy
  merger, BH feedback has little effect on the total number of stars
  formed, but is crucial for setting the BHs mass.
\end{abstract}

\section{Introduction}

Modern theories of galaxy formation hold that strong feedback
processes regulate star formation in galaxies across a wide range of
masses.  For more massive galaxies, stellar feedback processes appear
to become less efficient and feedback from a central massive black
hole (BH) begins to dominate. Feedback from an active galactic nucleus
(AGN) has been invoked to account for many observational results in
galaxy formation, including the $M_{BH} - M_{*}$ and $M_{BH} - \sigma$
relations and the suppression of star formation in elliptical galaxies
\citep{1998A&A...331L...1S,2003ApJ...596L..27K,2005ApJ...618..569M,2005Natur.433..604D,springel2005b,2007ApJ...669...67H}.

Many recent studies developing numerical models for the
effects of BHs on galactic scales have used broadly similar implementations
of the uncertain physics of AGN fueling and feedback (e.g.,
\citealt{2005MNRAS.361..776S, kawata2005, 2009ApJ...690..802J}).  It is,
e.g., often assumed that a BH of mass $M_{BH}$ accretes mass from the
surrounding interstellar medium (ISM) at a rate proportional to the Bondi rate
\citep{1952MNRAS.112..195B},
\begin{equation}
\dot{M}_{Bondi} = \frac{4 \pi f G^2 M_{BH}^2 \rho}{c_s^3},
\label{mdotbondieqn}
\end{equation}
\noindent where $\rho$ is the surrounding gas density, $c_s$ is the
sound speed and $f \sim 10-100$ takes into account the fact that the 
sphere of influence of the black hole is not always resolved 
\citep{2009MNRAS.398...53B}.  Moreover, these same calculations assume 
that the black hole's impact on its host galaxy can be approximated 
by depositing thermal energy released by accretion back into the 
surrounding gas.  There is, however, little detailed motivation or 
justification for either of these assumptions.

Eq.~(\ref{mdotbondieqn}) is only applicable when the gas fueling the
central BH has very little angular momentum.  Otherwise, the transport
of angular momentum regulates the accretion rate onto the central BH
(e.g., \citealt{shlosman1990}).  It is generally believed that
gas-rich disk galaxies are the progenitors of today's $\gtrsim L^*$
ellipticals and, in particular, that mergers of gas-rich galaxies lead
to luminous starbursts and the growth of central massive BHs
\citep{1988ApJ...325...74S,2005ApJ...630..705H}.  Most of the gas in
disk galaxies, merging galaxies, luminous starbursts
\citep{1998ApJ...507..615D,2006ApJ...640..228T}, and nearby luminous
AGN \citep{ho2008} appears to reside in a rotationally supported disk.
There is therefore no strong reason to believe that the
spherically-symmetric Bondi accretion rate is a reasonable estimate of
the accretion rate onto a BH in gas-rich disk galaxies.

The energy generated by a central AGN can couple to its surroundings
in a variety of ways, all of which may have a significant dynamical
influence on gas in the host galaxy and in the surrounding
intergalactic medium.  For example, relativistic jets inject energy
into intracluster plasma and may be the key mechanism for suppressing
cooling flows in galaxy clusters \citep{mcnamara2007}, although the
details of how the energy in the jet couples to the surrounding plasma
in a volume-filling way are not fully understood \citep{vernaleo2006}.
On galactic scales, winds from an accretion disk around the BH may
sweep-up and drive gas out of galaxies (e.g.,
\citealt{2003ApJ...596L..27K}).  And the AGN's radiation can strongly
impact the surrounding gas, both via Compton cooling/heating (e.g.,
\citealt{2005MNRAS.358..168S}) and via the momentum imparted as UV radiation is
absorbed by dust grains \citep{chang1987,1988ApJ...325...74S}.

The precise physical mechanism(s) responsible for AGN feedback are not
fully understood, particularly on galactic scales.  For this reason,
it is useful to distinguish between two classes of models: energy and
momentum injection.  
We believe that momentum injection, not energy injection, is likely
the dominant form of feedback for the majority of the gas in a galaxy.
In most circumstances, jets take the path of least resistance and
travel relatively unimpeded out of a galaxy. Furthermore, while
radiation from an AGN can, in principle, Compton heat the surrounding
gas enough to unbind it, it can only do so for very low density gas.
For example, for a BH radiating at $\sim 10^{46}$ erg $\rm{s}^{-1}$
with a typical quasar spectrum, only gas with $n \lesssim 1 \,
\rm{cm}^{-3}$ can be heated to the Compton temperature within $\sim
100$ pc.  However, the mean gas densities in the central $\sim 100 \,
\rm{pc}$ of ultraluminous infrared galaxies are $\sim 10^4
\rm{cm}^{-3}$ \citep{1998ApJ...507..615D}.  At these densities, the
cooling time of the gas is very short and the gas is unable to retain
any injected energy. Thus 
if the radiation from a BH strongly modifies the dynamics of the gas 
in its immediate vicinity, it must be via the {\it force} exerted when 
the radiation is absorbed.

Given the uncertainties in the physics of BH accretion and feedback,
it is important to explore a range of models for the impact of BHs on
galaxy formation.  Towards this end, we have carried out numerical
simulations of major galaxy mergers, qualitatively taking into account
the physics of accretion induced by angular momentum transport and AGN
feedback by momentum injection (radiation pressure).  Our accretion
and feedback prescriptions both differ from those used in previous
numerical simulations of BH growth and feedback.  The results in this
{\it Letter} are taken from a larger set of calculations (DeBuhr
et. al. in prep.) and represent general features of all the
simulations we have carried out.

\vspace{-0.5cm}
\section{Methods}

We use a non-public update of the TreeSPH code GADGET-2
\citep{2005MNRAS.364.1105S} to perform simulations of galaxy mergers
with feedback from both star formation and central supermassive BHs.
The code, provided by V. Springel, includes the effective star
formation model of \cite{2003MNRAS.339..289S}.  We describe below the
additional modifications that we have implemented to model BH growth
and feedback.

The multiphase equation of state of \cite{2003MNRAS.339..289S}
overpredicts the ``sound speed'' as compared to observations of the
random velocities in galaxies (in atomic or molecular gas).  For
example, the parameter choices $T_{SN}=4 \times 10^8 \rm{K}$, $A_0 =
4000$, $t_{*}^0 = 8.4 \rm{Gyr}$, and $q_{EOS} = 0.5$, which have been
used in previous works \citep{2005MNRAS.361..776S}, predicts $dv \sim
30$ km $\rm{s}^{-1}$ at $n \sim 1 \, \rm{cm}^{-3}$ and $dv \sim 110$
km $\rm{s}^{-1}$ for $n \sim \, 10^3 \rm{cm}^{-3}$.  These are too
large by a factor of $\sim 2-3$ compared with the observed values
\citep{1998ApJ...507..615D}.  To account for this difference, we
reduce the pressure everywhere by a constant factor of $10$.  As in
\cite{2003MNRAS.339..289S}, we assume that $\dot{\rho}_{*} \sim
\rho^{1.5}$ for consistency with observations
\citep{1988ApJ...334..144K}.  The normalization of the star formation
prescription is chosen so that a Milky Way-like galaxy has a total
star formation rate of about one solar mass per year; for galaxies
with different surface densities, the result is also consistent with
\cite{1988ApJ...334..144K}.  By reducing the pressure at fixed $\rho$
by a factor of 10, the gas is more dense in hydrostatic equilibrium.
This would increase the star formation rate relative to the observed
value.  To correct for this, we modify the equation of state
parameters to: $t_0^{*} = 13.86$ Gyr, $\beta = 0.1$, $A_0 = 6600$,
$T_{SN} = 6.6 \times 10^8$ K, $T_c = 1000$ K and $q_{EOS} = 0.5$.

The simulations described in this work are all mergers of equal mass
galaxies.  Each model galaxy consists of a dark matter halo, a
rotationally supported disk of gas and stars, a stellar bulge, and a
central BH.  The galaxy parameters are similar to those in
\cite{2005MNRAS.361..776S}: each galaxy has a total mass of
$1.36\times10^{12} M_{\sun}$; the mass of the disk is $4.1 \%$ of the
total, i.e., $5.57\times10^{10} M_{\sun}$, where 10\% of the disk mass
is assigned to gas and 90\% to stars; the bulge has a mass of
$1.86\times10^{10} M_{\sun}$, i.e., 1/3 of the total disk mass.  Each
galaxy is made of $8.0\times 10^5$ simulation particles and the
gravitational force softening is $\epsilon = 47$ pc.  The halo and the
bulge have \cite{1990ApJ...356..359H} density profiles, where the
virial and half-mass radii of the halo are 229 kpc and 102 kpc,
respectively (the concentration is 9.0), and the effective radius of
the bulge is 1.27 kpc.  The gaseous and stellar disks have exponential
profiles with scale lengths $R_d = 3.51$ kpc; the scale-height of the
stellar disk is $0.7$ kpc, while the scale-height of the gaseous disk
is determined by hydrostatic equilibrium.  The massive BH in each
galaxy is modeled using a specially marked collisionless tracer
particle.

The initial conditions are generated as in \cite{2005MNRAS.361..776S},
except for the decrease in gas pressure described above.  The galaxies are
placed on a prograde parabolic orbit.  The individual spins of the two
galaxies are randomly chosen to have a relative angle of about $41$
degrees.  The galaxies begin at a distance of $142$ kpc and the orbit has a
pericenter of $14.2$ kpc.

We estimate the accretion rate onto the BH from the surrounding gas,
due to viscous transport of angular momentum, using 
\begin{equation}
\dot{M}_{vis} = 3 \pi \alpha \Sigma \frac{c_s^2}{\Omega} \,,
\label{mdotvisceqn}
\end{equation}  
where $\Sigma$ is the mean surface density of the gas in the disk,
$\Omega$ is its angular rotational frequency, and the free parameter
$\alpha$ is the dimensionless viscosity; $\dot M_{vis}$ is also capped
at the Eddington rate.  We compute $\Sigma$ and the sound speed $c_s$
by taking an average of the properties of the individual SPH particles
in a spherical region with a radius $R_{acc} = 4 \epsilon = 188$ pc
centred on the BH.  Using velocity information directly from the
simulation particles themselves to compute $\Omega$ proved to be too
noisy; we thus determined $\Omega$ using the total mass, $M_T$, inside
$R_{acc}$ through $\Omega^2 = G M_T / R_{acc}^3$.  

Note that in our accretion prescription, $\dot M_{vis} = 0$ if there
is no gas within $4 \epsilon$ of the BH.  This feature of our model
accounts for the fact that our simulations capture the angular
momentum transport produced by gravitational torques on large scales
($\gtrsim 4 \epsilon = R_{acc}$); we assume, as is physically
reasonable but by no means proven, that these must be sufficient to
bring gas close to the BH (within $R_{acc}$) for significant BH
accretion to proceed.

Eq.~(\ref{mdotvisceqn}) is reminiscent of an alpha prescription of
\cite{1973A&A....24..337S}, but in
this formulation $\alpha$ parameterises both the efficiency of angular
momentum transport on scales smaller than our gravitational force
softening (by, e.g., gravitational torques) and the uncertainty due to
the fraction of the inflowing mass that turns into stars vs. accreting
onto the central BH.  
The physical processes responsible for transporting gas from $\sim$ kpc 
to $\sim 0.1$ pc are still not fully understood \citep{2003MNRAS.339..937G}, 
but non-axisymmetric gravitational torques are likely responsible 
\citep{1989Natur.338...45S}.
Detailed calculations of the structure
of AGN disks from $\sim 0.01-100$ pc, based on transport by spiral waves,
show that
Eq.~(\ref{mdotvisceqn}), evaluated at radii $\sim 30-100$ pc, can
provide a reasonable estimate of the accretion rate onto the BH in
some cases \citep{2005ApJ...630..167T}.
Although Eq.~(\ref{mdotvisceqn}) is only a crude
approximation to the true accretion rate onto the BH, it captures the
qualitative physics of accretion induced by angular momentum
transport, and is thus, we believe, a more suitable ``subgrid'' model
than Eq.~(\ref{mdotbondieqn}).
Our fiducial choice for $\alpha$ is
$\alpha = 0.05$, but we also present results for 0.15 in Sec.~3.
In the future, we intend to better calibrate our model of angular momentum
transport using simulations that focus on the central $\sim 100$ pc of
galaxies \citep{2009arXiv0912.3257H}.
 
To model the feedback onto the gas surrounding the BH, we have implemented
a simple phenomenological model in which the AGN's luminosity $L$ is coupled
back into the surrounding gas by depositing momentum radially outward from the BH.
Our goal is to account for the radiation pressure produced by the
absorption and scattering of the AGN's radiation by dust in the ISM.
To accurately do so would require a radiation transport calculation,
which is beyond the scope of the current paper.  Instead,
we model the impact of this radiation on the surrounding galaxy
by depositing a total momentum (per unit time) of 
\begin{equation}
 \tau L/c \,, \quad {\rm where} \quad L = {\rm min}(\eta \dot{M}_{vis} c^2, L_{Edd}) 
 \label{momdepeqn}
\end{equation}
radially away from the BH into every SPH particle within a distance of
$R_{acc}$ from the BH; 
each particle
receives the same acceleration.  
Note that the number of particles that receive this extra force, $N$,
will change with time as particles enter and leave the central region of 
radius $R_{acc}$.
We assume a radiative efficiency of $\eta
= 0.1$. 
Equation~(\ref{momdepeqn}) models the absorption by the dust of the UV radiation from the AGN 
(one $L/c$), and, more importantly, the subsequent diffusion of the
far IR photons ($\tau L/c$). In this way, the
value of $\tau$ determines the total momentum deposited and
corresponds to the total far IR optical depth in the nuclear region; we choose
$\tau = 10$ in these calculations.
This value of
$\tau$ is consistent with the fact that even the far-infrared
radiation produced by dust is optically thick at radii $\sim 100$ pc during
galaxy mergers (e.g., \citealt{2005ApJ...630..167T}).  
In particular, $\tau \sim 10$ is motivated by the surface density of
$\Sigma \sim 3$ g cm$^{-2}$ in the inner $\sim$ 100 pc near the peak of
accretion, and a FIR opacity of $\sim 3$ cm$^2$ g$^{-1}$.
The exact value of
$\tau$ does not significantly affect our conclusions, but it does normalize
the values of $\dot M_{vis}$ and $M_{BH}$ (see Eq.~\ref{mcriteqn} below).

The strength of the feedback on an individual particle depends not only on the
luminosity, but on the number of particles, $N$, to which the force is being 
applied in a given timestep. Our results do not depend strongly on $N$; this is
because the momentum is quickly shared with the rest of the gas particles via
pressure forces.  We carried out a number of test problems on the evolution of
gaseous shells with the additional force $\tau L / c$; these explicitly
show no dependence on $N$ (DeBuhr et. al. in prep). 

Computing the accretion rate onto and the feedback from the BH in the
simulations is prone to noise induced by the stochastic motion of the
BH particle.  To avoid this ``Brownian'' motion, we choose a mass for
the BH tracer particle of $2.8 \times 10^7 M_{\sun}$, which is roughly
a factor of $100$ higher than the other particle masses in the
simulation.  Note that this mass is an artificial dynamical mass for
simulation purposes; in addition to this, we integrate $\dot M(t)$ to
determine the ``true'' $M_{BH}$.  Once the two BH tracers have a
separation of $4 R_{acc}$ or smaller, we consider that
they would coalesce to form a single BH.
Once the tracers merge, the two values of $M_{BH}$ are summed and
one of the BH particles is moved to the center of mass of the two
tracers, and the other is removed from the region.

\vspace{-0.5cm}
\section{Results}

\begin{figure}
\includegraphics[width=84mm]{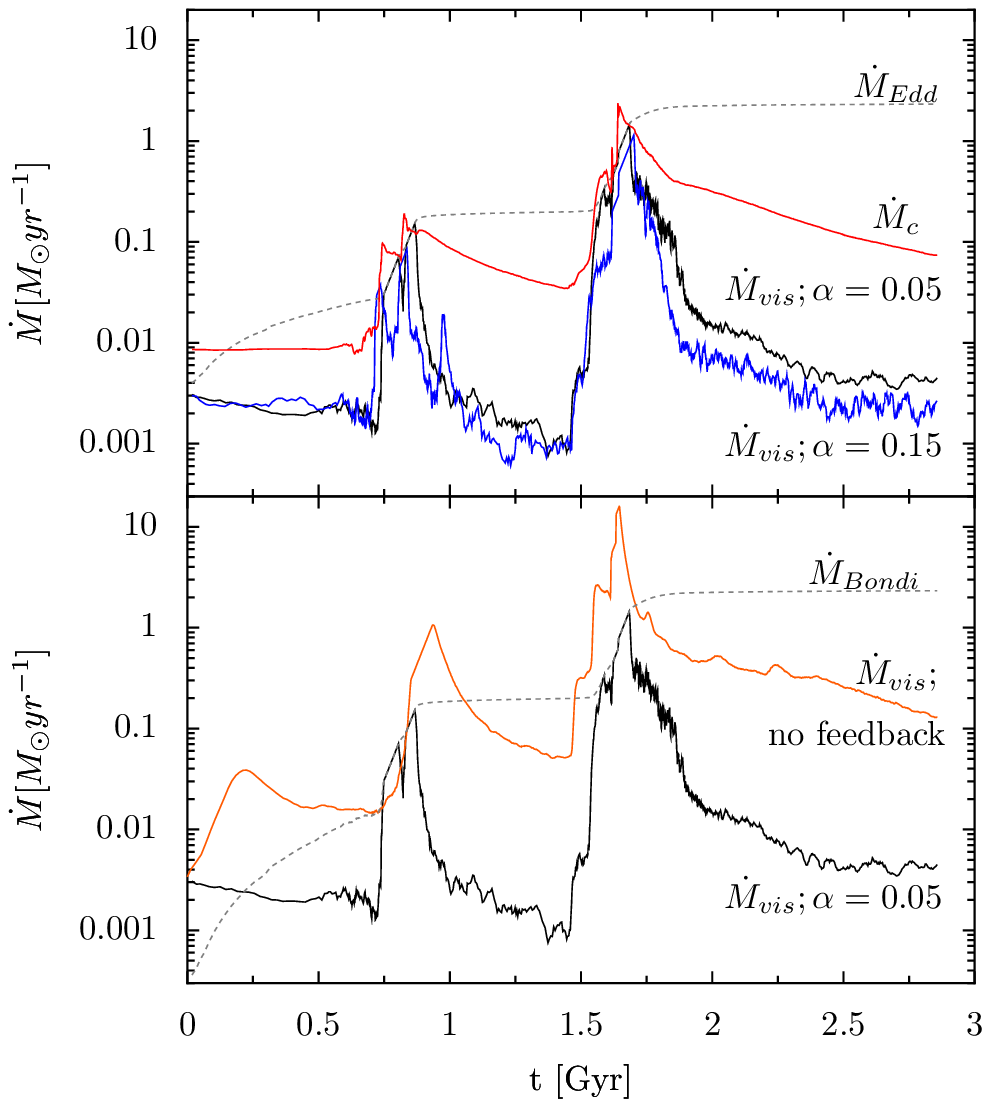}
\caption{{\it Top panel:} Viscous accretion rate $\dot M_{vis}$ onto
  the black hole (eq. [\ref{mdotvisceqn}]) for our fiducial galaxy
  merger simulation ($\alpha = 0.05$; black), and for a run with three
  times the viscosity ($\alpha = 0.15$; blue). Also shown is the
  critical mass accretion rate $\dot M_c$ (eq.~[\ref{mcriteqn}]) at
  which radiation pressure can push gas out of the nuclear region
  (red), and the Eddington rate $\dot M_{Edd}$ (grey).  Note that the
  accretion rate adjusts to $\dot M_{vis} \sim \dot M_c$ during the
  peaks of activity, independent of $\alpha$.  {\it Bottom panel:}
  Viscous accretion rate for the fiducial simulation (black) as
  compared to a simulation without feedback (orange; $\alpha = 0.15$);
  also shown is the Bondi accretion rate using the BH mass from the
  fiducial simulation. Unlike $\dot M_{vis}$, the Bondi rate is
  $\simeq \dot M_{Edd}$ at nearly all times.}
\label{figureMdotAlphas}
\end{figure}

The top panel of Figure~\ref{figureMdotAlphas} shows the viscous
accretion rate onto the BH for the fiducial run with $\alpha = 0.05$
(black) and for a run with $\alpha = 0.15$ (blue).  For comparison,
the Eddington rate $\dot M_{Edd} \equiv L_{Edd}/\eta c^2$ is shown in
grey, using the BH mass as a function of time from our fiducial
simulation.  The accretion rate is relatively constant at early times
but then peaks during the first close passage of the two galaxies at
$\sim 0.75$ Gyr and then even more strongly as the two galaxies
complete their merger at $\sim 1.6$ Gyr; note that $\dot M \ll \dot
M_{Edd}$ at both early and late times but reaches $\sim \dot M_{Edd}$
for $\sim 100$ Myrs near both first and final passage.

One of the interesting results in Figure~\ref{figureMdotAlphas} is
that differences in $\alpha$ do not significantly change the accretion
rate onto the BH, particularly near the peaks of activity.  This is
contrary to what one might expect from the fact that $\dot M_{vis} \propto
\alpha$ (eq. [\ref{mdotvisceqn}]).  
The origin of the weak dependence of $\dot M_{vis}$ on $\alpha$ is that
when the supply of mass is large, feedback from accretion onto the BH
regulates the rate at which the BH accretes.  Previous work has shown
that there is a critical luminosity $L_c$ at which the outward
radiation pressure force due to the central AGN just balances the
inward force of gravity.  For a simple spherically symmetric problem,
this is given by $\tau L_c/c = 4 f_g \sigma^4 /G$,
where $f_g$ is the gas fraction in the nuclear region and $\sigma^2 = G M_t
/ 2 R_{acc}$, with $M_t$ the total mass within $R_{acc}$
\citep{2005ApJ...618..569M}.  This in turn implies a critical accretion
rate $\dot{M}_c$
\begin{equation}
  \dot{M}_c = \frac{4 f_g}{\tau \eta c G} \sigma^4.
\label{mcriteqn}
\end{equation}
For $L \gtrsim L_c$, the radiation force on gas in the nuclear region
exceeds the inward force of gravity, and thus gas in the vicinity of
the BH will be pushed out of the nuclear region.  In our model, the
accretion rate is determined by the gas properties {\it within}
$R_{acc} \simeq 188$ pc; thus if the gas is largely pushed out of the
nuclear region, the accretion rate onto the BH decreases.  When $L
\lesssim L_c$, gravitational torques can drive gas into the nuclear
region towards the BH, thus increasing $\dot M_{vis}$.  This suggests that
the accretion rate may self-adjust such that $\dot M_{vis} \sim \dot M_c$.
To quantify this, the top panel of Figure~\ref{figureMdotAlphas} shows
$\dot{M}_c$ computed within $2 R_{acc}$ of the BH for the fiducial
calculation (red).  The accretion rate is indeed $\sim \dot M_c$ near
the peaks of activity.  This highlights that although $\dot M_{vis} \ll
\dot M_c$ is certainly possible if there is insufficient gas in the
nuclear regions (e.g., after the merger), feedback limits the maximum
rate at which the BH can accrete to be $\sim \dot M_c$. 
One point that we return to below is that although feedback does have
a strong effect on the gas dynamics in the galactic nuclei, it is {\it
  not} strong enough to blow large amounts of gas out of the galaxy as
a whole.

The bottom panel of Figure~\ref{figureMdotAlphas} compares the
accretion rate for the fiducial run (black) and a similar run with no
BH feedback (orange; $\alpha = 0.15$).  The peak accretion rate is a
factor of $\sim 10$ higher in the case without feedback, and the
duration of activity is significantly longer; moreover, $\dot M_{vis}$ in
the absence of feedback is $\propto \alpha$ and so can be scaled up or
down by arbitrary amounts by varying $\alpha$, unlike in the presence
of feedback (top panel).  Also shown in the lower panel is the Bondi
accretion rate (grey) with the BH mass set by that in the fiducial run
(which uses $\dot M_{vis}$).  For nearly all of the simulation, $\dot M_{Bondi}
\simeq \dot M_{Edd}$, and unlike the simulations of
\cite{2005MNRAS.361..776S} or \cite{2009ApJ...690..802J}, the Bondi
rate does not decrease after the final merger, because the ambient gas
remains cool and dense.

\begin{figure}
\includegraphics[width=84mm]{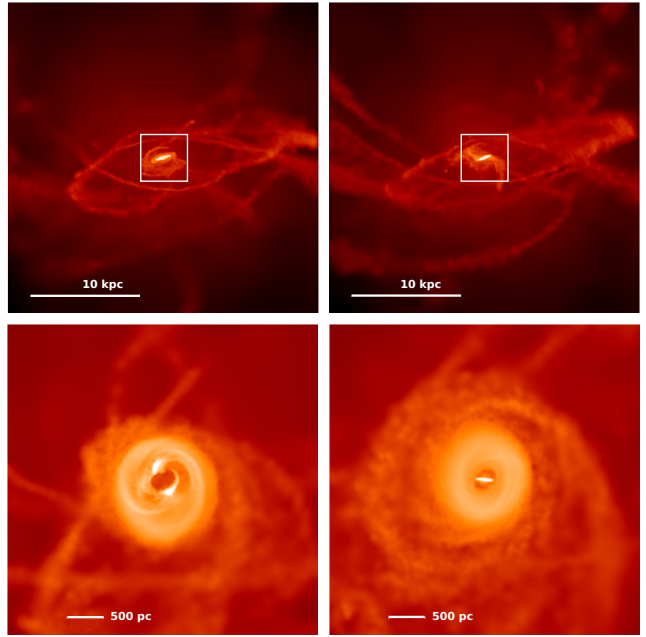}
\caption{Column densities of gas centered on the BH for the run with
  feedback (left) and without feedback (right), at $t = 1.74$ Gyr
  during the peak of activity (Fig.~\ref{figureMdotAlphas}).  
  The colour indicates the column density with brighter colour
  indicating larger densities.  The large scale images are projected onto
  the orbital plane, whereas the small scale images are viewed along the
  orbital plane as the nuclear disk has a large inclination.  The white
  box shows the relative scale of the images.
  Note that the large-scale gas
  distribution is quite similar in the two cases; the BHs only visible
  effect is in the nuclear region ($\lesssim 300$ pc), where some of
  the gas is cleared out in the case with feedback.  These images were
  generated using SPLASH \citep{2007PASA...24..159P}.}
\label{figureGasImage}
\end{figure}

Figure~\ref{figureGasImage} shows the column density of gas in the vicinity
of the BH for the case without feedback (right) and the case with feedback
(left) both at the same time during the peak of accretion at 
$t = 1.74 \rm{Gyr}$.
The images are $28.5 \rm{kpc}$ on a side in the top row, and $4.28
\rm{kpc}$ on a side in the bottom row.
In the simulation with feedback, one can see explicitly that the gas
has been evacuated from the region near the BH (within $\sim
R_{acc}$), as argued above.  These images also demonstrate that the
feedback from the BH does not produce a large-scale blow-out of matter
from the galactic nucleus.  More quantitatively, at the end of the
simulation, the runs with and without feedback have the same mass of
gas outside $4 R_d$ to within $10\%$, and the
mass that is at large radii is due to the merger dynamics (e.g., tidal
tails) rather than the BH driving a powerful outflow.  This is
qualitatively consistent with observational evidence for large
reservoirs of atomic and molecular gas in nearby luminous AGN and
quasars \citep{scoville2003}, which have relatively normal kinematics
\citep{ho2008}.  By contrast, previous simulations using the Bondi
accretion rate and thermal feedback find that the BH unbinds the
remaining gas in the galaxy near the end of the merger (e.g.,
\citealt{2005Natur.433..604D}) and that this can be important for
shutting off star formation in ellipticals \citep{springel2005b}.

\begin{figure}
\includegraphics[width=84mm]{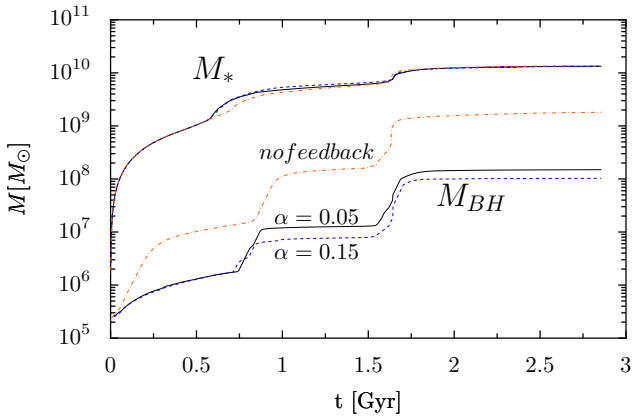}
\caption{The black hole mass $M_{BH}$ (lower three curves) as a function of
  time during the merger, and the mass of new stars formed during the
  merger $M_{*}$, for three simulations: runs including BH feedback with
  $\alpha = 0.05$ (black) and $\alpha = 0.15$ (blue, dashed), and the run 
  without BH feedback (orange, dash-dot).  Note that BH feedback has little 
  effect on the
  total stellar mass formed, while it reduces the final BH mass by a factor
  of $\sim 10$ compared to the run without feedback.  The BH mass is,
  however, nearly independent of $\alpha$ because the accretion
  self-regulates as shown in Figure \ref{figureMdotAlphas}.}
\label{figureMasses}
\end{figure}

In Figure~\ref{figureMasses} we show the BH mass $M_{BH}$ and the
integrated mass of stars formed during the merger ($M_{*}$) for the
fiducial simulation (black), the run without feedback (orange, dash-dot), 
and the
run with $\alpha = 0.15$ (blue, dash).  The addition of feedback changes 
the total star formation during the merger by less than $1\%$.  
In addition, the final BH
masses for the two runs with feedback differ by about $30\%$.
This is a further consequence of the self regulated accretion during
most stages of the merger.  There is, of course, freedom in choosing
the initial mass of the BH in our calculations, but so long as this is
sufficiently small, it does not significantly change the final mass of
the BH.  For the simulations without feedback, the final BH mass is
larger than in the presence of feedback by a factor of $\simeq 10$, as
would be expected from Figure \ref{figureMdotAlphas}.  In addition,
the mass of the BH scales $\propto \alpha$ in this case.  Although we
have not made a quantitative comparison, the small dispersion in BH
mass for different subgrid accretion models in the presence of
feedback appears consistent with the small scatter in the
$M_{BH}-\sigma$ relation (while models without feedback would produce
a larger dispersion in $M_{BH}$).

\vspace{-0.5cm}
\section{Discussion}

The model presented here is a necessarily simplified treatment of the
physics occurring in the nuclear regions of galaxies.  In particular,
the choices of $\alpha$ in the accretion model and of $R_{acc}$ and $\tau$
in the model of momentum deposition are somewhat uncertain.  Changing
these values affects some of details of the gas dynamics.  For
instance, the accretion history of the BH change modestly as we vary
$R_{acc}$ and $\tau$.  However, it is encouraging that many of the global
results of the simulations are insensitive to these choices.  The peak
luminosity occurs at the same time and always reaches the Eddington
limit.  The total stellar mass formed during a merger is essentially
independent of these parameters and the final BH mass is relatively
insensitive to both $R_{acc}$ and $\alpha$.  Perhaps most interestingly, the
peak accretion rate is relatively independent of the subgrid accretion
model ($\alpha$) and is instead set by the structure of the host
galaxy and the feedback physics, reaching the critical rate $\sim \dot
M_c$ at which radiation pressure balances gravity in the nuclear
regions of the galaxy (eq.~\ref{mcriteqn}).

A clear next step in this modeling effort is to perform radiative
transfer calculations simultaneously with the SPH calculation in order
to more reliably determine the radiation pressure force.  This would
not only eliminate the need to specify the parameters $R_{acc}$ and $\tau$
by hand, but would also provide information about the AGN spectrum as
a function of time. Detailed comparisons between these results and
observations should allow quantitative tests of the importance of AGN
feedback by momentum deposition during BH growth and galaxy formation.

\vspace{-0.5cm}
\section*{Acknowledgments}
\vspace{-0.15cm} We thank Volker Springel for providing the version of
Gadget used here.  JD and EQ were supported in part by NASA grant
NNG06GI68G and the David and Lucile Packard Foundation. Support for
PFH and EQ was provided in part by the Miller Institute for Basic
Research in Science, University of California Berkeley.
This research used resources of the National Energy Research Scientific 
Computing Center, which is supported by the Office of Science of the U.S.
Department of Energy under Contract No. DE-AC02-05CH11231.
This work was partially supported by the National Center for
Supercomputing Applications under AST080048 and utilized the Intel 64
cluster Abe.

\vspace{-0.6cm}

\bibliographystyle{mn2e}
\bibliography{letter}

\end{document}